\documentclass[prd,twocolumn,nofootinbib,superscriptaddress,fleqn,amssymb,amsfonts,amsmath]{revtex4-2}
\usepackage{graphicx,slashed,xcolor,multirow,bbold,mathtools,sidecap,tikz,bm,ulem,enumitem,booktabs,array}
\usepackage[colorlinks=true,linktoc=page,linkcolor=purple,citecolor=teal,urlcolor=magenta]{hyperref}
\usepackage[compat=1.1.0]{tikz-feynman}
\graphicspath{ {figures/} }

\definecolor{lime}{HTML}{A6CE39}
\DeclareRobustCommand{\orcidicon}{\hspace{-2.1mm}
\begin{tikzpicture}
\draw[lime,fill=lime] (0,0.0) circle [radius=0.13] node[white] {{\fontfamily{qag}\selectfont \tiny ID}}; \draw[white,fill=white] (-0.0525,0.095) circle [radius=0.007]; 
\end{tikzpicture} \hspace{-3.7mm} }
\foreach \x in {A, ..., Z}
 {\expandafter\xdef\csname orcid\x\endcsname{\noexpand\href{https://orcid.org/\csname orcidauthor\x\endcsname} {\noexpand\orcidicon}}}

\setenumerate{label=$(\roman*)$,itemsep=1pt,parsep=1pt,topsep=1pt,partopsep=1pt}
\begin{document}

\title{Low-mass doubly-charged Higgs bosons at LHC}

\author{Saiyad Ashanujjaman\orcidA{}}
\email{saiyad.a@iopb.res.in}
\affiliation{Institute of Physics, Bhubaneswar, Sachivalaya Marg, Sainik School, Bhubaneswar 751005, India}                                                           
\affiliation{Department of Physics, SGTB Khalsa College, Delhi 110007, India}
\affiliation{Department of Physics and Astrophysics, University of Delhi, Delhi 110007, India}

\author{Kirtiman Ghosh}
\email{kirti.gh@gmail.com}
\affiliation{Institute of Physics, Bhubaneswar, Sachivalaya Marg, Sainik School, Bhubaneswar 751005, India}                                                           
\affiliation{Homi Bhabha National Institute, Training School Complex, Anushakti Nagar, Mumbai 400094, India}

\author{Rameswar Sahu\orcidC{}}
\email{rameswar.s@iopb.res.in}
\affiliation{Institute of Physics, Bhubaneswar, Sachivalaya Marg, Sainik School, Bhubaneswar 751005, India}                                                           
\affiliation{Homi Bhabha National Institute, Training School Complex, Anushakti Nagar, Mumbai 400094, India}
                                                                                                                                                                                      
\begin{abstract}
Search for light (within the mass range 84–200 GeV) doubly-charged Higgs bosons decaying into a pair of W-bosons has been deemed challenging using the conventional LHC searches with leptons, jets and missing transverse momentum in the final state. Such Higgses together with slightly heavier singly-charged and neutral Higgses, when arranged in an $SU(2)_L$ triplet as in the type-II see-saw model, are lately shown to accommodate the recent measurement of the $W$-boson mass by the CDF collaboration. These, when produced in a highly Lorentz-boosted regime, tend to manifest themselves as a single fat-jet or a pair of adjacent same-sign leptons plus missing transverse momentum. First, we perform a multivariate analysis to discern such exotic jets from the SM jets. Then, we present a novel search in the final state with an exotic jet and two same-sign leptons plus missing transverse momentum. We find that such low-mass doubly-charged Higgsses could be directly probed with the already collected Run 2 LHC data.
\end{abstract}

\maketitle                                                                                           

\section{\label{sec:intro} Introduction}
Despite being remarkably successful in understanding particle physics phenomenology, the Standard Model (SM) in its present form lacks a mass term for the neutrinos. However, a trivial Dirac mass term for the neutrinos can be effectuated by dint of the usual Higgs mechanism by introducing right-handed neutrinos to the SM. Although plausible, this warrants philosophical displeasure as it calls for diminutive Yukawa couplings. Conversely, a well-founded remedy to this menace is offered by the so-called {\it see-saw mechanism}, wherein a lepton number violating New Physics beyond the SM is invoked at {\it a priori} unknown scale---presumably away from both the electroweak (EW) scale and the Planck scale, so that on integrating out the heavy fields, the SM neutrinos are left with observed sub-eV masses after the EW symmetry breaking. Pointedly, numerous models of varying complexity and testability at colliders have been proposed over the last few decades. The type-II see-saw model \cite{Konetschny:1977bn,Cheng:1980qt,Lazarides:1980nt,Schechter:1980gr,Mohapatra:1980yp,Magg:1980ut}, a UV completion of the Weinberg operator at the tree level \cite{Weinberg:1979sa,Ma:1998dn}, extending the SM with an $SU(2)_L$ triplet scalar field with hypercharge $Y=1$, is arguably the most widely-studied variant \cite{Huitu:1996su,Gunion:1996pq,Chakrabarti:1998qy,Chun:2003ej,Muhlleitner:2003me,Akeroyd:2005gt,Akeroyd:2007zv,Garayoa:2007fw,Han:2007bk,Kadastik:2007yd,delAguila:2008cj,FileviezPerez:2008jbu,Perez:2008ha,Akeroyd:2009hb,Akeroyd:2010ip,Arhrib:2011uy,Melfo:2011nx,Aoki:2011pz,Akeroyd:2011zza,Arbabifar:2012bd,Chiang:2012dk,Akeroyd:2012nd,Chun:2012jw,Chun:2012zu,delAguila:2013mia,Chun:2013vma,Kanemura:2013vxa,Dev:2013ff,Kanemura:2014goa,Kanemura:2014ipa,kang:2014jia,Deppisch:2015qwa,Han:2015hba,Han:2015sca,Blunier:2016peh,Das:2016bir,Mitra:2016wpr,Cai:2017mow,Ghosh:2017pxl,Nomura:2017abh,Antusch:2018svb,BhupalDev:2018tox,Crivellin:2018ahj,Agrawal:2018pci,Rahili:2019ixf,deMelo:2019asm,Dev:2019hev,Primulando:2019evb,Chun:2019hce,Padhan:2019jlc,Bandyopadhyay:2020mnp,Ashanujjaman:2021txz,Yang:2021skb,Ashanujjaman:2022tdn}. For one, the flavour structure of the Yukawa coupling driving the leptonic decays of the triplet-like scalars ensues to be governed by the neutrino oscillation data up to the scalar triplet VEV. Moreover, the presence of the doubly-charged scalars ($H^{\pm \pm}$) and their characteristic decays to a pair of same-sign leptons ($\ell^\pm \ell^\pm$) or $W$-bosons offer interesting ways to probe them directly at the current and near-future experiments. 

The experimental collaborations have carried out several searches for $H^{\pm \pm}$ \cite{ATLAS:2012hi,Chatrchyan:2012ya,ATLAS:2014kca,Khachatryan:2014sta,CMS:2016cpz,CMS:2017pet,Aaboud:2017qph,CMS:2017fhs,Aaboud:2018qcu,Aad:2021lzu,ATLAS:2021jol,ATLAS:2022yzd}, and non-observations of any significant excess over the SM expectations have led to stringent limits on them. For $H^{\pm \pm}$ decaying into $\ell^\pm \ell^\pm$, the ATLAS collaboration has set a lower limit of 1020 GeV assuming equal branching fractions across modes \cite{ATLAS:2022yzd}. This search considers only light leptons in the final states, and thus not sensitive for $H^{\pm \pm}$ decaying into $\tau^\pm \tau^\pm$. The CMS collaboration has set a lower limit of 535 GeV on such scalars \cite{CMS:2017pet}. For $H^{\pm \pm}$ decaying into $W^\pm W^\pm$, the ATLAS collaboration has excluded them within the mass range 200--350 GeV considering their Drell-Yan pair production \cite{ATLAS:2021jol}. An orderly re-interpretation of this search considering all possible Drell-Yan production modes for the triplet-like scalars results in an improved exclusion range of 200--400 GeV \cite{Ashanujjaman:2021txz}. Moreover, a re-interpretation of the ATLAS same-sign dilepton search in Ref.~\citep{ATLAS:2014kca} has derived an exclusion limit of 84 GeV \cite{Kanemura:2014ipa}.

In a nutshell, $H^{\pm \pm}$ decaying into $WW^{(*)}$ are still allowed in the 84--200 GeV mass window. In this mass window, the type-II see-saw model predicts a cross-section between 1.5 pb to 65 fb for $pp \to H^{++} H^{--}$ at the 13 TeV LHC. Despite a sizeable cross-section, searching such an $H^{\pm \pm}$ using the conventional LHC searches with leptons, jets, and missing transverse momentum in the final state has been challenging. the CMS and ATLAS collaborations have turned a blind eye to this. Presumably, for one, their eventual decay products tend to be not so hard and are likely to be drowned in the LHC environment owing to the inherent towering EW and QCD backgrounds. Moreover, ineludible contamination from the SM resonances makes the state of affairs worse. To the extent of our knowledge, the only notable effort in probing this mass window was made in Ref.~\cite{Kang:2014lwn}. Lately, Refs.~\cite{Kanemura:2022ahw,Heeck:2022fvl,Bahl:2022gqg,Cheng:2022hbo} have demonstrated that the recently reported measurement of the $W$-bosoon mass by the CDF experiment \cite{CDF:2022hxs} which substantially differs from the global EW fit \cite{Awramik:2003rn} can be explained within the type-II see-saw model predicting such low-mass $H^{\pm \pm}$ and slightly heavier singly-charged and neutral scalars. Therefore, it is paramount to look for such $H^{\pm \pm}$ at the LHC.

In this work, we present a novel search strategy for such $H^{\pm \pm}$. We consider their pair production in a highly Lorentz-boosted regime such that they are produced back-to-back with large transverse momenta, manifesting themselves as a single fat-jet or a pair of adjacent same-sign leptons plus missing transverse momentum. Obviously, this would reduce the signal cross-section significantly. However, should we be able to discern such exotic jets from the SM jets, a final state with such a jet and two same-sign leptons plus missing transverse momentum would have a compensating advantage of reducing the SM background more aggressively, thereby ameliorating the signal-to-background ratio. Keeping that in mind, first, we perform a multivariate analysis incorporating the jet mass, jet charge, $N$-subjettiness, {\it etc.} variables as inputs to the boosted decision tree (BDT) classifier to discern such exotic jets (dubbed $H^{\pm\pm}$-jets hereafter) from the SM jets. Then, we perform a search in the final state with an $H^{\pm\pm}$-jet and two same-sign leptons plus missing transverse momentum. 

The rest of this work is structured as follows. In Section~\ref{sec:model}, we briefly discuss the doubly-charged Higgses in the type-II see-saw model. We perform a detailed collider analysis in Section~\ref{sec:collider}. Finally, we summarise in Section~\ref{sec:conclusion}.

\section{\label{sec:model} The doubly-charged Higgses}
In the type-II see-saw model, the SM is augmented with an $SU(2)_L$ triplet scalar field with hypercharge $Y=1$
\[
\Delta = \begin{pmatrix} \Delta^+/\sqrt{2} & \Delta^{++} \\ \Delta^0 & -\Delta^+/\sqrt{2} \end{pmatrix}.
\]
The scalar potential involving $\Delta$ and the SM Higgs doublet $\Phi = \begin{pmatrix} \Phi^+ \!&\! \Phi^0 \end{pmatrix}^T $ is given by
\begin{align*}
V(\Phi,\Delta) &= -m_\Phi^2{\Phi^\dagger \Phi} + \frac{\lambda}{4}(\Phi^\dagger \Phi)^2 + m_\Delta^2{\rm Tr}(\Delta^{\dagger}{\Delta}) 
\\
& + [\mu(\Phi^T{i}\sigma^2\Delta^\dagger \Phi)+{\rm h.c.}] + \lambda_1(\Phi^\dagger \Phi){\rm Tr}(\Delta^{\dagger}{\Delta})
\\
&  + \lambda_2[{\rm Tr}(\Delta^{\dagger}{\Delta})]^2 + \lambda_3{\rm Tr}[(\Delta^{\dagger}{\Delta})^2] + \lambda_4{\Phi^\dagger \Delta \Delta^\dagger \Phi},
\end{align*}
where $m_\Phi^2, m_\Delta^2$ and $\mu$ are the mass parameters, $\lambda$ and $\lambda_i$ ($i\!=\!1,\dots,4$) are the dimensionless quartic couplings, and $\sigma^2$ is one of the Pauli matrices. The neutral components $\Phi^0$ and $\Delta^0$ procures respective VEVs $v_d$ and $v_t$ that $\sqrt{v_d^2+2v_t^2}=246$ GeV. For detailed discussions of the main dynamical features of the scalar potential, see Refs.~\cite{Arhrib:2011uy,Arbabifar:2012bd,Chun:2012jw,Das:2016bir}. After the EW symmetry is broken, the degrees of freedom carrying identical electric charges mix, thereby resulting in several physical Higgs states: 
\begin{enumerate}
\item the neutral states $\Phi^0$ and $\Delta^0$ mix into two CP-even states ($h$ and $H0$) and two CP-odd states ($G^0$ and $A^0$), 
\item the singly-charged states $\Phi^\pm$ and $\Delta^\pm$ mix into two mass states $G^\pm$ and $H^\pm$, 
\item the doubly-charged state $\Delta^{\pm \pm}$ is aligned with its mass state $H^{\pm \pm}$.
\end{enumerate}

\noindent The mass states $G^0$ and $G^\pm$ are the {\it would-be} Nambu-Goldstone bosons, $h^0$ is identified as the 125 GeV Higgs observed at the LHC, and the rest follows the sum rule
\[
m_{H^{\pm\pm}}^2-m_{H^\pm}^2 \approx m_{H^\pm}^2 - m_{H^0/A^0}^2 \approx -\frac{\lambda_4}{4}v_d^2.
\]

The Yukawa interaction $Y^{\nu}_{ij} L^T_i C i \sigma^2 \Delta L_j$ ($L_i$ stands for the SM lepton doublet with $i\in e,\mu,\tau$, and $C$ the charge-conjugation operator) induces masses for the neutrinos:
\[
m_\nu=\sqrt{2}Y^\nu v_t.
\]

\begin{figure}[htb!]
\centering
\includegraphics[width=0.9\columnwidth]{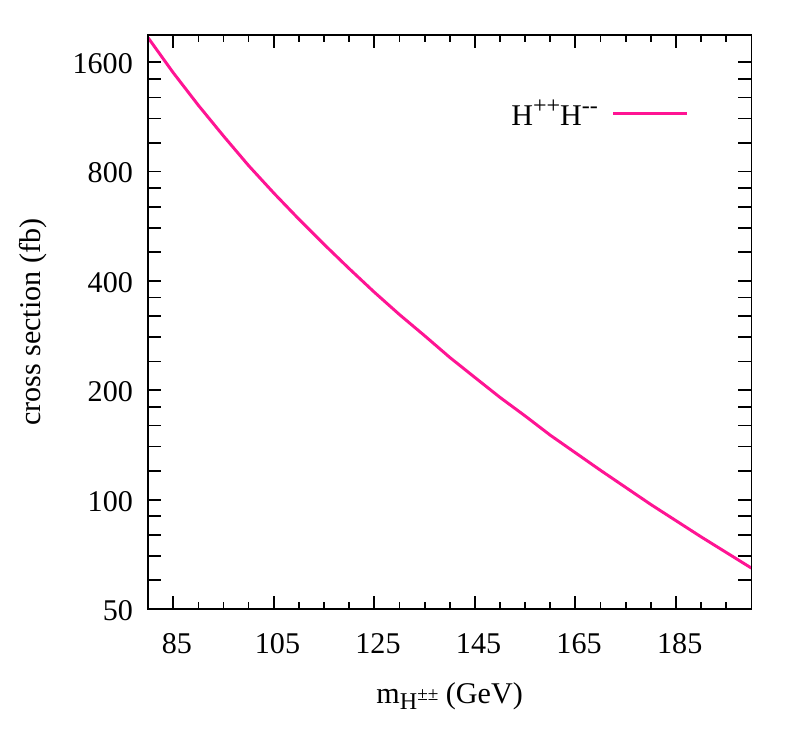}
\caption{\label{fig:xsec}LO cross-section for $pp \to H^{++} H^{--}$ at the 13 TeV LHC.}
\end{figure}

The doubly-charged Higgses are pair produced aplenty at the LHC by quark-antiquark annihilation via the neutral current Drell-Yan mechanism:\footnote{They are also produced via $t/u$-channel photon fusion as well as vector-boson fusion processes. However, such processes are rather sub-dominant.}
\[
q \bar{q} \to \gamma^*/Z^* \to H^{++} H^{--}.
\]
We evaluate the leading order (LO) cross-sections using the \texttt{SARAH 4.14.4} \cite{Staub:2013tta,Staub:2015kfa} generated \texttt{UFO} \cite{Degrande:2011ua} modules in \texttt{MadGraph5\_aMC\_v2.7.3} \cite{Alwall:2011uj,Alwall:2014hca} with the \texttt{NNPDF23\_lo\_as\_0130\_qed} parton distribution function \cite{Ball:2013hta,NNPDF:2014otw}. Fig.~\ref{fig:xsec} shows the LO doubly-charged Higgs pair production cross-section at the 13 TeV LHC as a function of their mass. Following the relevant QCD corrections estimated in Refs.~\cite{Muhlleitner:2003me,Fuks:2019clu}, we naively scale the LO cross-section by an overall next-to-leading (NLO) $K$-factor of 1.15. Therefore, the resulting $pp \to H^{++}H^{--}$ cross-section varies from 1.72 pb to 74.5 fb for 84 GeV to 200 GeV mass.

After being produced, $H^{\pm \pm}$ decays into $\ell^\pm \ell^\pm$, $W^\pm W^{\pm(*)}$ and $H^\pm W^{\pm *}$, if kinematically allowed. In broad terms, the dominance of one decay mode over the others depends on three parameters, namely $m_{H^{\pm \pm}}$, $v_t$ and $\Delta m=m_{H^{\pm \pm}}-m_{H^\pm}$, see Refs.~\cite{FileviezPerez:2008jbu,Aoki:2011pz,Ashanujjaman:2021txz} for detailed discussions. For the present work, without commiting to a fixed value for $v_t$ and $\Delta m$, we assume exclusive prompt decays of $H^{\pm \pm}$ to $W^\pm W^{\pm(*)}$.

\section{\label{sec:collider} Collider analysis}
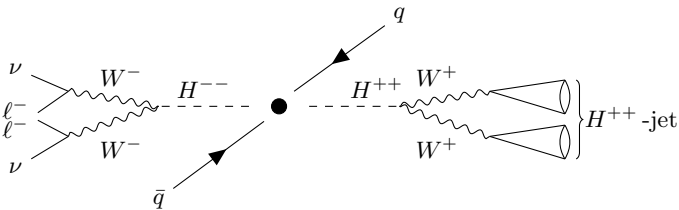
\begin{figure}[htb!]
\begin{tikzpicture}
\begin{feynman}
\vertex[large, dot] (c) {};
\vertex[right=0.4cm of c] (cr);
\vertex[left=0.4cm of c] (cl);
\vertex[right=1.2cm of cr] (r1);
\vertex[right=1.2cm of r1] (r20);
\vertex[above=0.2cm of r20] (r2);
\vertex[below=0.4cm of r20] (r3);
\vertex[right=1.0cm of r2] (r40);
\vertex[above=0.15cm of r40] (r4);
\vertex[below=0.3cm of r40] (r5);
\vertex[right=1.0cm of r3] (r60);
\vertex[above=0.15cm of r60] (r6);
\vertex[below=0.3cm of r60] (r7);
\vertex[left=1.2cm of cl] (l1);
\vertex[left=1.2cm of l1] (l20);
\vertex[above=0.2cm of l20] (l2);
\vertex[below=0.4cm of l20] (l3);
\vertex[left=0.7cm of l2] (l40);
\vertex[above=0.12cm of l40] (l4) {$\nu$};
\vertex[below=0.24cm of l40] (l5) {$\ell^-$};
\vertex[left=0.7cm of l3] (l60);
\vertex[above=0.12cm of l60] (l6) {$\ell^-$};
\vertex[below=0.24cm of l60] (l7) {$\nu$};
\vertex[right=0.15cm of r4] (r41);
\vertex[right=0.15cm of r7] (r71);
\vertex[right=0.2cm of c] (u10);
\vertex[above=0.2cm of u10] (u1);
\vertex[above=1.0cm of r1] (u2) {$q$};
\vertex[left=0.2cm of c] (d10);
\vertex[below=0.2cm of d10] (d1);
\vertex[below=1.0cm of l1] (d2) {$\bar{q}$};
\diagram*{
(cr) -- [scalar, edge label=\(H^{++}\), near end] (r1) -- [boson, edge label=\(W^+\), near end] (r2),
(r2) -- (r4), (r2) -- (r5), (r4) -- [bend left] (r5) -- [bend left] (r4),
(r3) -- (r6), (r3) -- (r7), (r6) -- [bend left] (r7) -- [bend left] (r6),
(r1) -- [boson, edge label'=\(W^+\), near end] (r3),
(cl) -- [scalar, edge label'=\(H^{--}\)] (l1) -- [boson, edge label'=\(W^-\), near end] (l2),
(l2) -- (l4), (l2) -- (l5),
(l3) -- (l6), (l3) -- (l7),
(l1) -- [boson, edge label=\(W^-\), near end] (l3),
(u2) -- [fermion] (u1),
(d2) -- [fermion] (d1)
};
\draw [decoration={brace}, decorate] (r41.north west) -- (r71.south west) node [pos=0.5, right] {\(H^{++}\operatorname{-jet}\)};
\end{feynman}
\end{tikzpicture}
\caption{\label{fig:Feynman} Schematic Feynman diagram for $q\bar{q} \to H^{++}H^{--}$ and its subsequent decays to one $H^{\pm \pm}$-jet, two same-sign leptons and neutrinos.}
\end{figure}

In this section, we present a novel search strategy for $H^{\pm\pm}$ with $m_{H^{\pm\pm}} \in$ [84--200] GeV. We only consider $H^{\pm\pm}$ which are produced in a highly Lorentz-boosted regime, manifesting themselves as a single fat-jet or a pair of adjacent same-sign leptons plus missing transverse momentum. Such a requirement significantly reduces the signal cross-section.\footnote{For example, a parton level cut of $p_T(H^{\pm\pm}) > 300$ GeV reduces the $pp \to H^{++}H^{--}$ cross-section by a factor of 48(4.4) to 37.4(17.0) fb for $m_{H^{\pm \pm}}=84(200)$ GeV.} As argued earlier, despite such a notable reduction in the signal cross-section, the final state with an $H^{\pm\pm}$-jet and two same-sign leptons plus missing transverse momentum (see Fig.~\ref{fig:Feynman}) is expected to have a compensating advantage of reducing the SM background more aggressively with the proviso that we discern the $H^{\pm\pm}$-jets from the SM jets.

In the following, we briefly describe the reconstruction and selection of various physics objects, then perform a multivariate analysis to discern the $H^{\pm\pm}$-jets from the SM jets, {\it viz.} QCD jets, $W/Z$-jets, $h$-jets, and $t$-jets, and finally delineate a search in the final state with an $H^{\pm\pm}$-jet and two same-sign leptons plus missing transverse momentum.

\subsection{\label{sec:object} Object reconstruction and selection}

We pass the parton-level events into {\tt PYTHIA~8.2} \cite{Sjostrand:2014zea} to simulate subsequent decays for the unstable particles, initial and final state radiations (ISR and FSR), showering, fragmentation and hadronisation, and then into {\tt Delphes~3.4.2} with the default CMS card \cite{deFavereau:2013fsa} for simulating detector effects as well as reconstructing various physics objects, {\it viz.} photons, electrons, muons and jets.

Constituents of the {\it would-be} fat-jets are clustered using the {\it anti-k$_T$ algorithm} \cite{Cacciari:2008gp} with a characteristic jet radius $R=1.0$ as implemented in {\tt FastJet 3.3.2} \cite{Cacciari:2011ma}. To remove the soft yet wide-angle QCD emissions from the fat-jets, we use the {\it jet pruning} algorithm \cite{Ellis:2009su,Ellis:2009me} with the default values for the pruning parameters: $z_{cut} = 0.1$ and $R_{cut} = 0.5$ \cite{Ellis:2009su}. Further, to unfold the multi-prong nature of the fat-jets, we use an inclusive jet shape termed as {\it $N$-subjettiness} $\tau_N$ \cite{Thaler:2010tr,Thaler:2011gf}\footnote{It is defined as $\tau_{N} = \frac{1}{d_0} \sum_k p_{T,k} {\rm min}\left(\Delta R^\beta_{1,k}, \Delta R^\beta_{2,k},...,\Delta R^\beta_{N,k} \right)$, where $N$ is the number of subjets a jet is presumably composed of, $k$ runs over the jet constituents with transverse momentum $p_{T,k}$, $\Delta R_{i,k}$ is the distance in the rapidity-azimuth plane between a candidate subjet $i$ and a jet constituent $k$, $d_0=\sum_k p_{T,k} R_0^\beta$ with $R_0(=1.0)$ being the characteristic jet radius used in the original jet clustering algorithm, and $\beta$ is an angular weighting exponent dubbed {\it thurst parameter}.} choosing {\it one-pass $k_T$-axes} for the minimisation procedure and $\beta=1$. Reconstructed jets are required to be within the pseudorapidity range $|\eta|<2.5$ and have a transverse momentum $p_T > 30$ GeV, whereas the leptons (electrons and muons) are required to have $|\eta|<2.5$ and $p_T > 10$ GeV. Moreover, we demand the scalar sum of the $p_T$s of all other objects lying within a cone of radius 0.3(0.4) around an electron (a muon) to be smaller than 10\%(15\%) of its $p_T$. This ensures that the leptons are isolated. Finally, the missing transverse momentum $\vec p_T^{\rm\,\,miss}$ (with magnitude $p_T^{\rm miss}$) is estimated from the momentum imbalance in the transverse direction associated to all reconstructed objects in an event.

\subsection{\label{sec:bdt} Multivariate analysis: discerning the $H^{\pm\pm}$-jets from the SM jets}
Here we perform a multivariate analysis with the BDT classifier implemented in the {\tt TMVA~4.3} toolkit integrated into the analysis framework {\tt ROOT~6.24}. For training and testing the classifier, we use 600000 events for each category of the SM jets and 300000 for each $m_{H^{\pm\pm}}$ within the [85,195] GeV range in steps of 10 GeV. Of these, 80\% are picked randomly for training, and the rest are used for testing.

We use the following kinematic features of the jets as inputs to the BDT classifier:
\begin{enumerate}
\item invariant mass $m$
\item $b$-tag\footnote{It is a {\it boolean} indicating whether or not at least one of the constituet subjet is a $b$-jet.}
\item jet charge $Q_k$ \cite{Krohn:2012fg}\footnote{Jet charge is defined as $ Q_{k}=\frac{\sum_i q_i \left(p_{T,i}\right)^k}{\sum_i p_{T,i}}$, where $i$ runs over the associated tracks with transverse momentum $p_{T,i}$ and charge $q_i$, and $k$ is a free regularisation exponent which we take to be 0.2.}
\item $N$-subjettiness variables $\tau_1,\tau_{21},\tau_{32}$ and $\tau_{43}$.\footnote{$\tau_{N,N-1}=\tau_N/\tau_{N-1}$ is an useful discriminant between $N$- and $(N-1)$-prong jets.}
\end{enumerate}

\begin{widetext}

\begin{figure}[htb!]
\centering
\includegraphics[width=0.32\columnwidth]{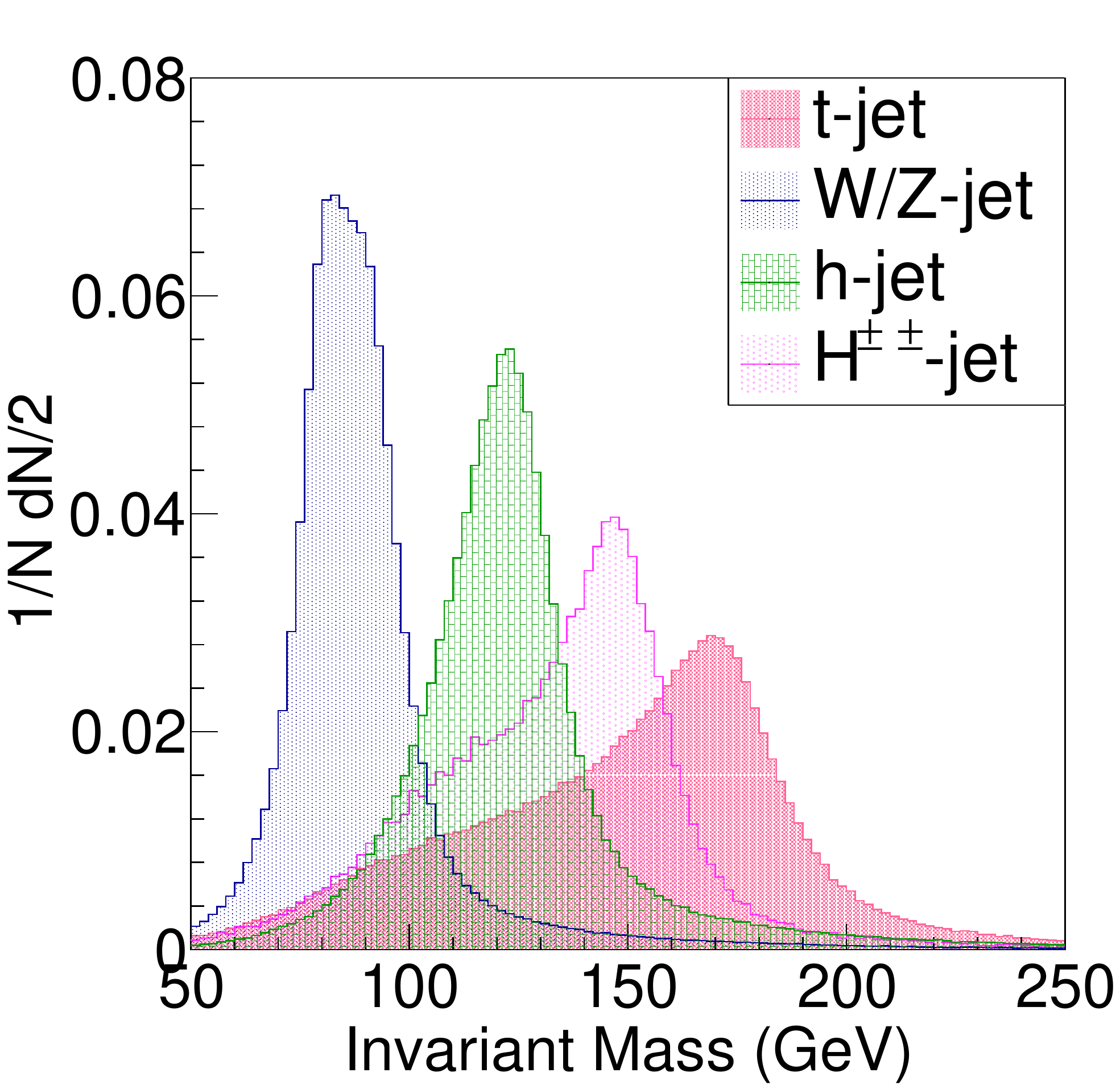}
\includegraphics[width=0.32\columnwidth]{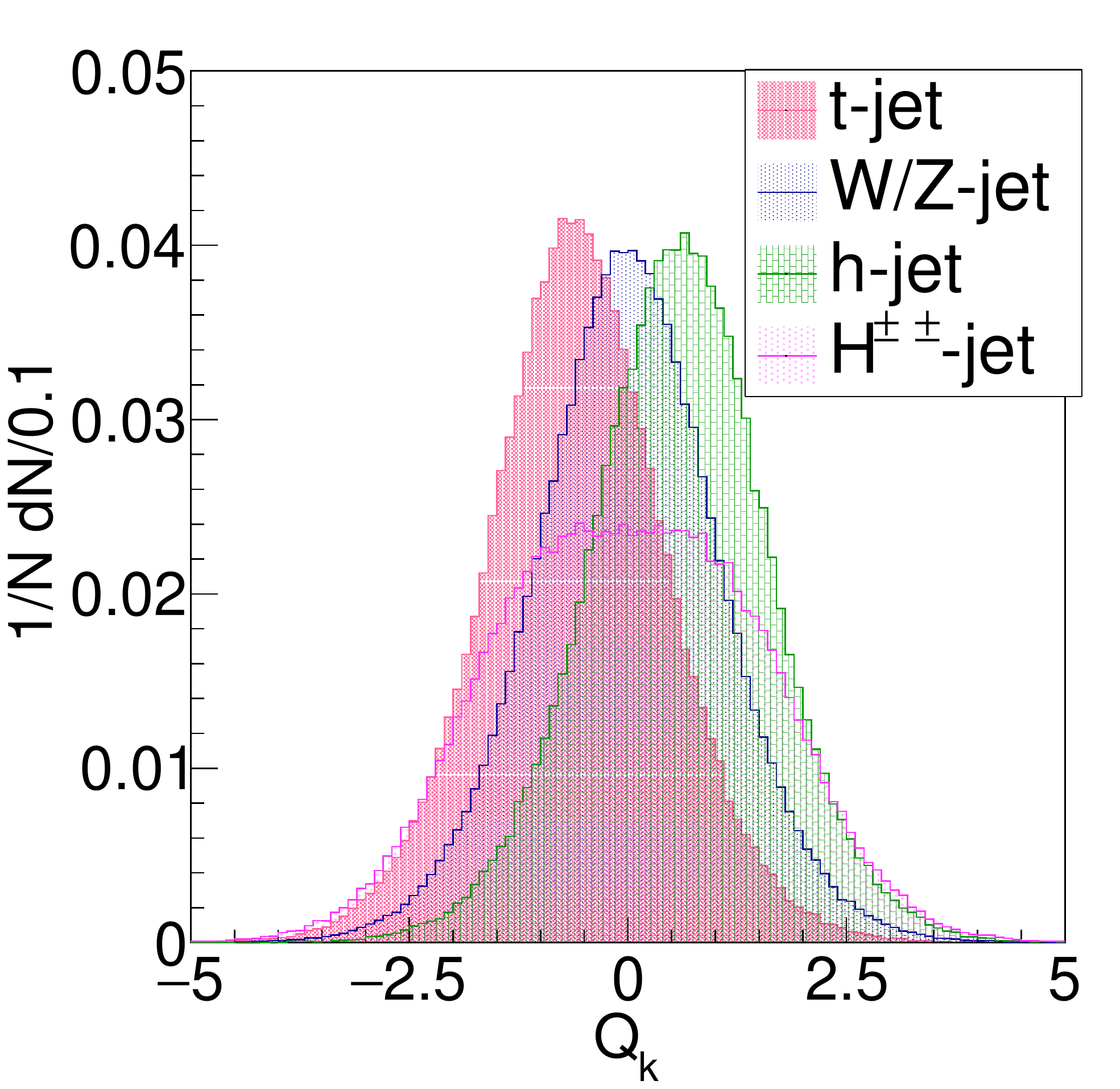}
\includegraphics[width=0.32\columnwidth]{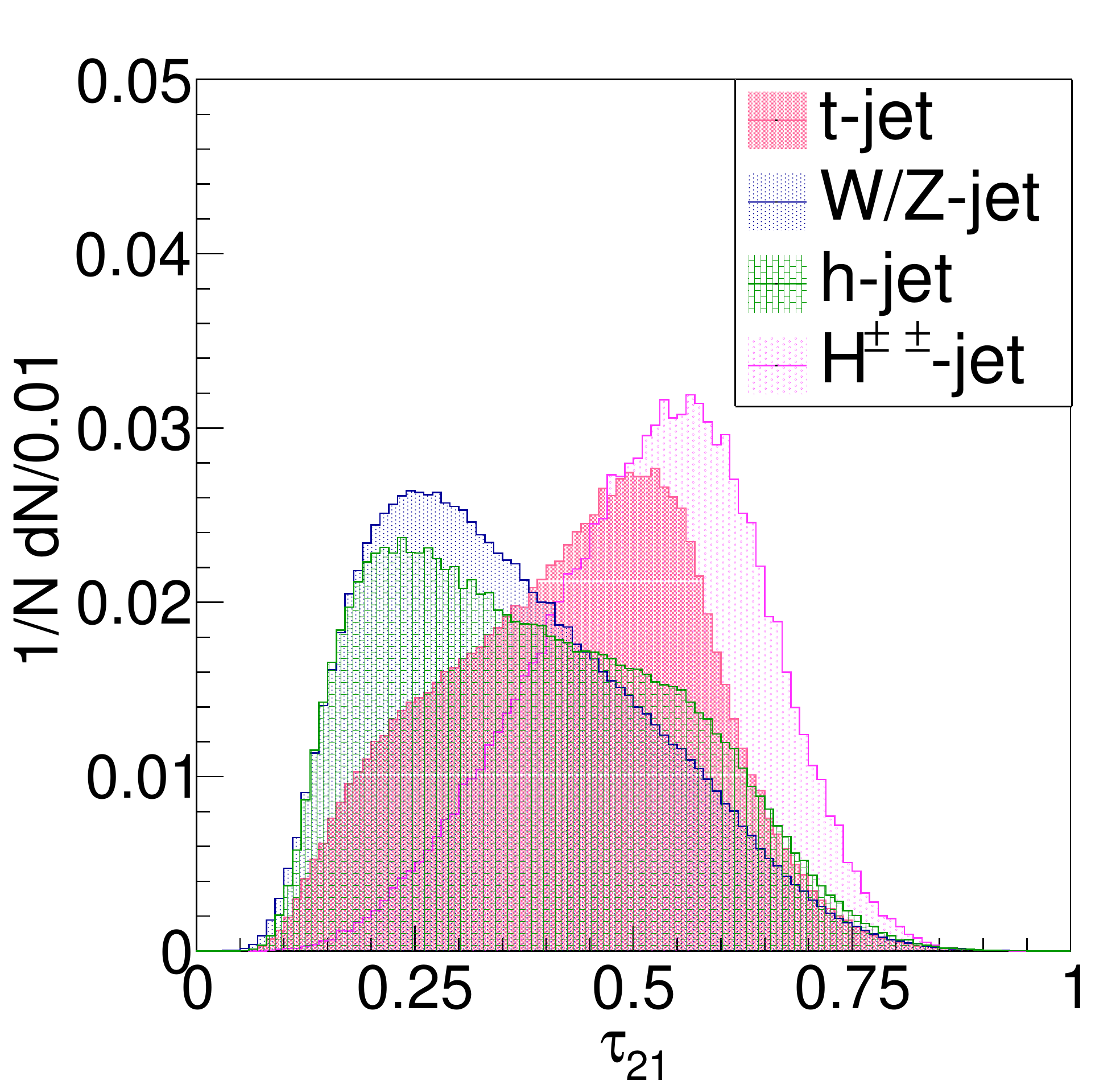}
\caption{\label{fig:bdtInput} Normalised distributions for some of the input features. The signal distributions are for $m_{H^{\pm\pm}}=150$ GeV.}
\end{figure}

\end{widetext}

\begin{figure}[htb!]
\centering
\includegraphics[width=0.49\columnwidth]{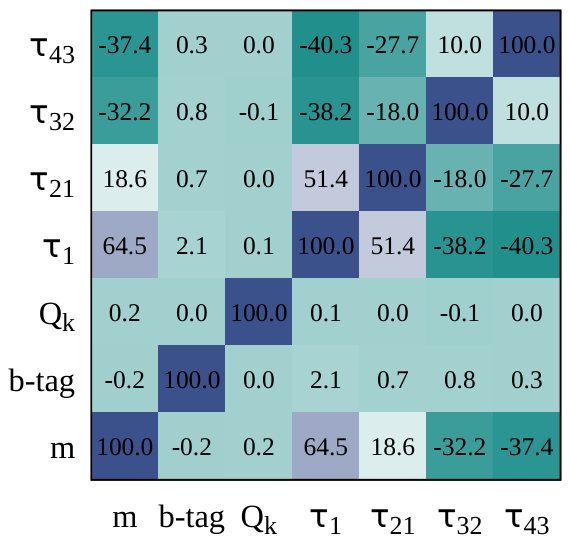}
\includegraphics[width=0.423\columnwidth]{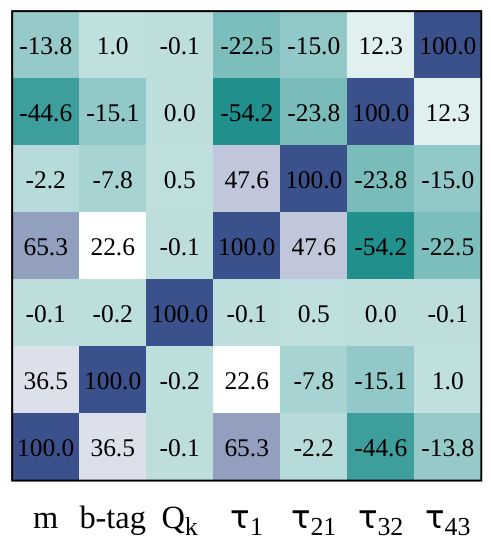}
\caption{\label{fig:bdtCorr} Correlations in \% among the input features for the $H^{\pm\pm}$-jets (left) and the SM jets (right).}
\end{figure}

\noindent The normalised distributions for some of the input features are shown in Fig.~\ref{fig:bdtInput}, the rest are not shown for brevity. These variables constitute a minimal set with $(a)$ good discrimination power between the $H^{\pm\pm}$-jets and the SM jets, and $(b)$ low correlations among themselves. The method-unspecific separation is a good measure of the former. For a given feature $x$, this is defined as
\[
\langle S^2 \rangle = \frac{1}{2} \int \frac{\left[\hat{x}_H(x)-\hat{x}_{SM}(x)\right]^2}{\hat{x}_H(x)+\hat{x}_{SM}(x)} dx
\]
where $\hat{x}_H(x)$ and $\hat{x}_{SM}(x)$ are the probability density functions of $x$ for the $H^{\pm\pm}$-jets and the SM jets, respectively. Table~\ref{table:bdtSeparationRanking} shows method-unspecific separation for the input features, while Fig.~\ref{fig:bdtCorr} show their Pearson’s linear correlation coefficients defined as
\[
\rho(x,y) = \frac{\langle xy \rangle - \langle x \rangle \langle y \rangle}{\sigma_x \sigma_y},
\]
where $\langle x \rangle$ and $\sigma_x$, respectively, are the expectation value and standard deviation of $x$.

\begin{table}
\centering
\begin{tabular}{lll}
\toprule
Feature & Method-unspecific & Method-specific \\ 
& separation & ranking \\ 
\midrule 
$m$ & 0.064 & 0.152 \\ 
$b$-tag & 0.099 & 0.167 \\
$Q_k$ & 0.052 & 0.101 \\ 
$\tau_1$ & 0.134 & 0.151 \\ 
$\tau_{21}$ & 0.104 & 0.208 \\ 
$\tau_{32}$ & 0.075 & 0.120 \\ 
$\tau_{43}$ & 0.066 & 0.102 \\ 
\bottomrule
\end{tabular} 
\caption{\label{table:bdtSeparationRanking} Method-unspecific separation and method-specific ranking of the input features.}

\end{table}

\begin{figure}[htb!]
\centering
\includegraphics[width=0.9\columnwidth]{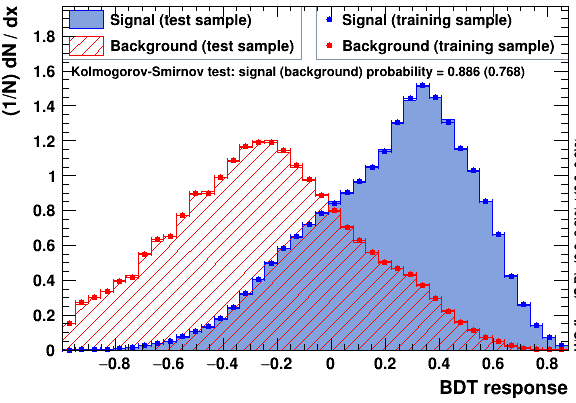}
\caption{\label{fig:bdtKS} BDT response curves for the training and testing subsamples.}
\end{figure}

To enhance the BDT classification, we use the {\it adaptive boost} algorithm with a learning rate of 0.1, and combine 1000 decision trees with 5\% minimum node size and a depth of 4 layers per tree into a forest. As the separation criterion for node splitting, we use the so-called {\it Gini index}. The relevant BDT hyperparameters are summarised in Table~\ref{table:bdtParametrs}. Table~\ref{table:bdtSeparationRanking} also shows the method-specific ranking of the input features. In other words, this shows the relative importance of the input features in separating the $H^{\pm\pm}$-jets from the SM jets. As we see from Table~\ref{table:bdtSeparationRanking}, the $N$-subjettiness variable $\tau_{21}$ is the best separating variable, while the jet-charge $Q_k$ is the one with least separating power. Finally, we check the classifier for overtraining by performing the Kolmogorov-Smirnov (KS) test which compares the BDT response curves for the training and testing subsamples, see Fig.~\ref{fig:bdtKS}. These response curves exhibit no considerable overtraining. 

\begin{table}[htb!]
\centering
\begin{tabular}{ll}
\toprule
BDT hyperparameter & Optimised choice \\
\midrule
NTrees & 1000 \\ 
MinNodeSize  & 5\% \\ 
MaxDepth & 4 \\ 
BoostType & AdaBoost\\ 
AdaBoostBeta & 0.1  \\ 
UseBaggedBoost & True \\ 
BaggedSampleFraction & 0.5 \\
SeparationType & GiniIndex \\ 
nCuts & -1 \\ 
\bottomrule
\end{tabular} 
\caption{\label{table:bdtParametrs} Summary of optimised BDT hyperparameters.}
\end{table}

\begin{figure}[htb!]
\centering
\includegraphics[width=0.49\columnwidth]{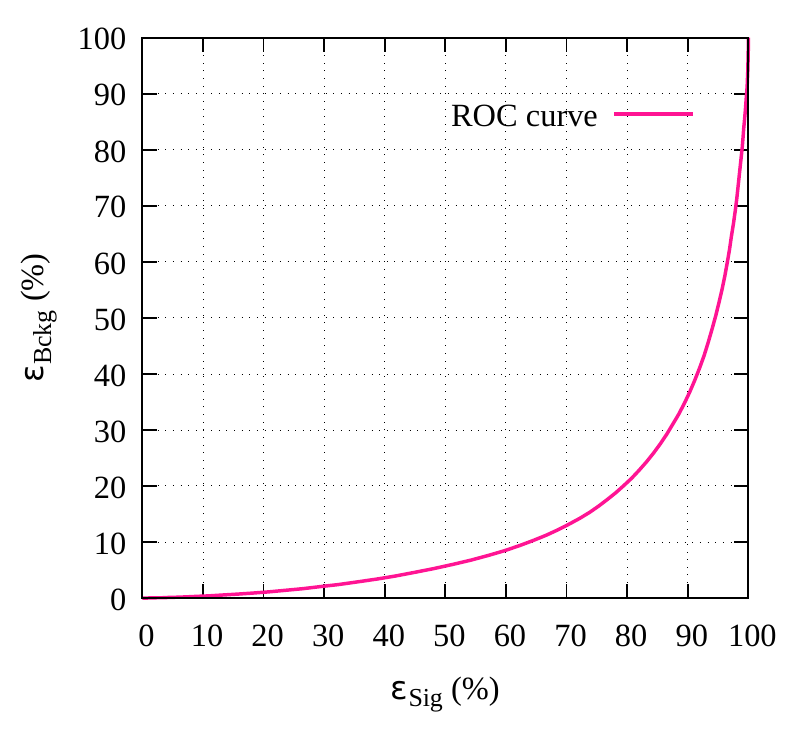}
\includegraphics[width=0.49\columnwidth]{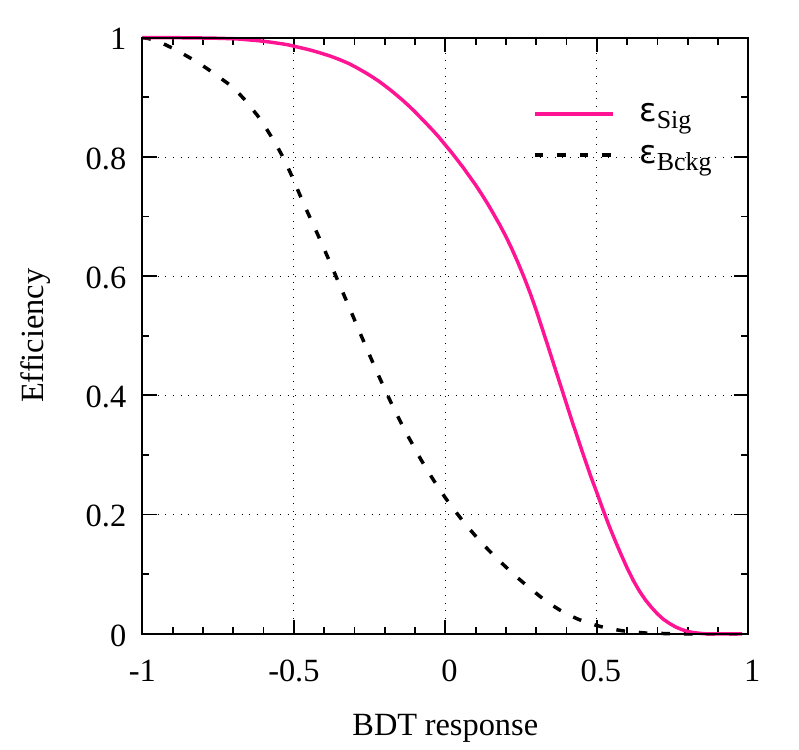}
\caption{\label{fig:ROC} Combined BDT performance in terms of the ROC curve (left), and the signal (with $m_{H^{\pm\pm}}=150$ GeV) and background efficiencies as a function of the BDT response.}
\end{figure}

\begin{figure}[htb!]
\centering
\includegraphics[width=0.99\columnwidth]{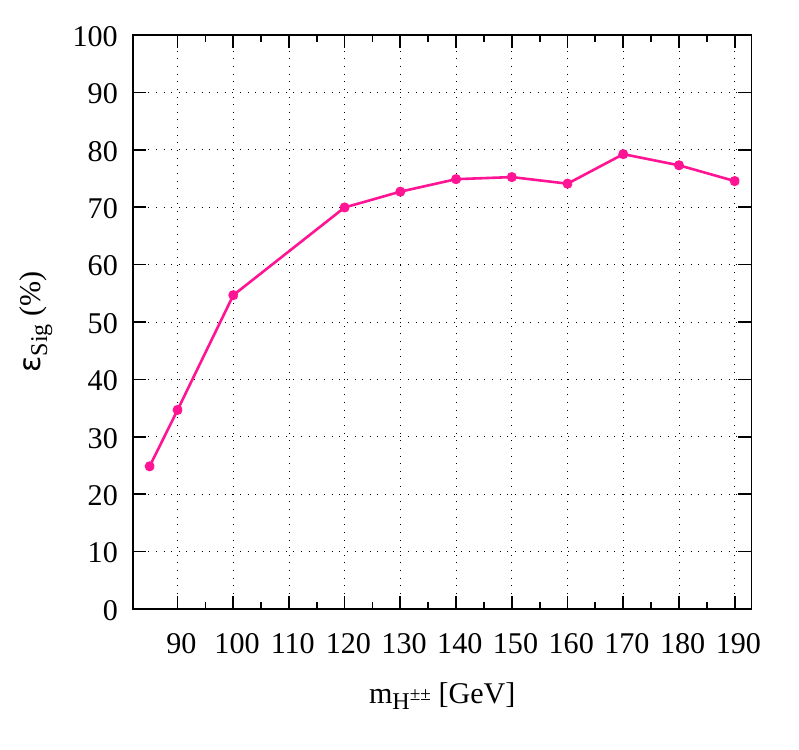}
\caption{\label{fig:effS} The signal efficiency as a function of $m_{H^{\pm\pm}}$ for the BDT response of 0.1.}
\end{figure}

In the left panel of Fig.~\ref{fig:ROC}, we show the receiver-operator-characteristic (ROC) curve, which quantifies the combined BDT performance, for $m_{H^{\pm\pm}}=150$ GeV. The right panel of Fig.~\ref{fig:ROC} shows the signal (with $m_{H^{\pm\pm}}=150$ GeV) and background efficiencies ($\epsilon_{\rm Sig}$ and $\epsilon_{\rm Bckg}$) as a function of the BDT response. The area below the ROC curve is $\sim 0.13$, indicating considerably well separation between the signal and background. For a BDT response greater than 0, not only $\epsilon_{\rm Bckg}$ but also $\epsilon_{\rm Sig}$ falls to lower values, whereas for a BDT response less than 0, both rises to higher values. Therefore, we choose an optimum value of 0.1 for the BDT response. In Fig.~\ref{fig:effS}, we show the variation of $\epsilon_{\rm Sig}$ with $m_{H^{\pm\pm}}$ for the chosen value of the BDT response. The abrupt drop in $\epsilon_{\rm Sig}$ for $m_{H^{\pm\pm}} \lesssim 100$ GeV is ascribed to the small mass difference between $m_{H^{\pm\pm}}$ and the $W$-mass. For small mass difference, the decay products of the off-shell $W$-boson emanating from $H^{\pm\pm}$ tend to be very soft, and thus are not likely to pass the object reconstruction and selection criteria discussed in Section~\ref{sec:object}. As a consequence of this, the features of an $H^{\pm\pm}$-jet resemble to that of an SM jet, thereby making the former indiscernible from the latter.

\subsection{\label{sec:backg} SM backgrounds}
As the background for the present analysis, we consider numerous SM processes such as diboson, triboson and tetraboson processes, Higgsstrahlung processes, single and multi-top productions in association with/without gauge bosons, and Drell-Yan processes. All these processes are generated in association with up to two jets at the LO using {\tt MadGraph5\_aMC\_v2.7.3} \cite{Alwall:2011uj,Alwall:2014hca} at least of worth 3000 fb$^{-1}$ luminosity of data at the 13 TeV LHC, followed by the {\it MLM matching} using {\tt PYTHIA~8.2} \cite{Sjostrand:2014zea}, and then naively scaled by appropriate NLO (or higher, whichever is available in the literature) $K$-factors \cite{Catani:2009sm,Balossini:2009sa,Campbell:2011bn,Cascioli:2014yka,Campbell:2016jau,LHCHiggsCrossSectionWorkingGroup:2016ypw,Shen:2016ape,Nhung:2013jta,Shen:2015cwj,Wang:2016fvj,Alwall:2014hca,Frederix:2014hta,Kidonakis:2015nna,Muselli:2015kba,Broggio:2019ewu,Frederix:2017wme}.

The relevant backgrounds can be broadly classified into two classes: prompt and non-prompt. While most of these processes contribute to the former, only the processes where a jet is misidentified as a lepton or additional leptons originate from ISR/FSR photon conversions and in-flight heavy-flavour decays constitute the latter. Though the lepton isolation requirement (mentioned in Section~\ref{sec:object}) and the $b$-jet veto (mentioned later in Section~\ref{sec:cuts}) significantly subdue the latter, a considerable fraction of this still passes the object selection. The estimation of this contribution requires a data-driven approach, naemly the so-called {\it fake factor} method, which is beyond the realm of this work. We adopt a conservative approach, assuming a $p_T$-dependent probability of 0.1–0.3\% for a jet to be misidentified as a lepton \cite{ATLAS:2016iqc}. Further, to account for the electron charge misidentification due to their bremsstrahlung interactions with the inner detector material, all prompt electrons are naively corrected with a $p_T$- and $\eta$-dependent charge misidentification probability: $P(p_T,\eta)=\sigma(p_T) \times f(\eta)$, where $\sigma(p_T)$ and $f(\eta)$ ranges from 0.02 to 0.1 and 0.03 to 1, respectively \cite{ATLAS:2017xqs}.

\subsection{\label{sec:cuts} Event selection and analysis}
Here we discuss the selection criteria that are adept in ameliorating the signal-to-background ratio. Only the events satisfying the following selection cuts ({\it S0}) are considered for further analysis:
\begin{enumerate}
\item one fat-jet with $p_T > 300$ GeV,
\item two same-sign leptons,
\item the angular separation between the leptons $\Delta R_{\ell\ell} > 0.05$,
\item the dilepton invariant mass $m_{\ell\ell} > 1$ GeV as well as $m_{\ell\ell} \notin [3,3.2]$ GeV.
\end{enumerate}

\noindent The requirements $\Delta R_{\ell\ell}>$ 0.05 and $m_{\ell\ell}>$ 1 GeV vanquishes the background contributions from muon bremsstrahlung interactions as well as ISR/FSR photon conversions, and $m_{\ell\ell} \notin [3,3.2]$ GeV suppresses contributions from $J/\psi$ decays.

The events satisfying the {\it S0} cut are then fed to the trained BDT classifier described in Section~\ref{sec:bdt}. Following the discussion in Section~\ref{sec:bdt}, we impose a modest cut on the BDT response
\[
{\rm {\it S1}:~ BDT~response} > 0.1.
\]

Figure~\ref{fig:mlldRll} shows the normalised distributoin of $m_{\ell\ell}$ for the signal with $m_{H^{\pm\pm}}=150$ GeV and background events satisfying the {\it S1} cut. For the signal, it is a monotonically falling distribution with an end point near 120 GeV as ocassioned by the low mass of $H^{\pm\pm}$. On the contrary, the background boasts a peak at the $Z$-boson mass with the lion's share of the contributions accruing from $Z\to e^-e^+$ when one of the electrons charge get misidentified. To supress the $Z\to e^-e^+$ contribution, we require that
\[
{\rm {\it S2}}:~ m_{\ell \ell} < 80 {\rm ~GeV}.
\]

In the left panel of Fig.~\ref{fig:mlldRll}, displayed is the normalised distribution for $p_T^{\rm miss}$ suggesting that the signal looks much harder than the background. Therefore, a reasonably strong cut on $p_T^{\rm miss}$ would be helpful in curtailing the latter without impinging much on the former. In Fig.~\ref{fig:mlldRll}, also displayed are the distributions for the angular separtion between the two leptons ($\Delta R_{\ell\ell}$) and the azimuthal separation between the dilepton system and $p_T^{\rm miss}$ ($\Delta \phi (\ell \ell, p_T^{\rm miss})$). As we see, unlike the background, most of the signal events are contained within $\Delta R_{\ell\ell} \sim 1$ and $\Delta \phi (\ell \ell, p_T^{\rm miss}) \sim 1$ showing that, as we expect, the leptons and neutrinos emanating from highly Lorentz-boosted $H^{\pm\pm}$ are adjacent to each other. Guided by these distribution, we impose the following set of cuts:
\[
{\rm {\it S3}}:~ \Delta R_{\ell \ell} < 1.2,~ p_T^{\rm miss} > 80 {\rm ~GeV},~ \Delta \phi (\ell \ell, p_T^{\rm miss}) < 0.8.
\]

\begin{table}[htb!]
\begin{center}
\begin{tabular}{l|c|c|c|c}
\toprule 
Event sample & {\it S0} & {\it S1} & {\it S2} & {\it S3} \\ 
\toprule
$\gamma^*/Z^*$ & 11.49 & 2.432 & 0.154 & 0.004 \\
$t\bar{t}$ & 3.931 & 0.436 & 0.120 & 0.028 \\
$W^\pm Z$ & 3.238 & 0.784 & 0.216 & 0.057 \\
$t\bar{t}W^\pm$ & 2.461 & 0.311 & 0.084 & 0.018 \\
$W^\pm W^\pm jj$ & 1.992 & 0.480 & 0.107 & 0.023 \\
$W^\pm$ & 1.985 & 0.473 & 0.334 & 0.116 \\	
$W^\pm W^\pm W^\mp$ & 1.474 & 0.284 & 0.076 & 0.022 \\
Others & 3.579 & 0.598 & 0.168 & 0.046 \\
\hline
Total background & 30.15 & 5.798 & 1.259 & 0.314 \\ 
\hline
Signal: $m_{H^{\pm\pm}}=90$ GeV & 0.946 & 0.387 & 0.387 & 0.312 \\
Signal: $m_{H^{\pm\pm}}=120$ GeV & 1.087 & 0.735 & 0.731 & 0.586 \\
Signal: $m_{H^{\pm\pm}}=150$ GeV & 0.976 & 0.652 & 0.560 & 0.434 \\ 
\bottomrule
\end{tabular} 
\caption{\label{table:cut-flow} Signal and background cross-sections (fb) after different selection cuts.}
\end{center}
\end{table}

Table~\ref{table:cut-flow} shows the progression of the background and signal (with $m_{H^{\pm\pm}}=90,120$ and 150 GeV) cross-sections at the 13 TeV LHC as subsequent selection cuts are imposed. As we see, all these cuts turn out be very efficacious in subjugating the background while keeping the signal relatively less harmed.

\subsection{\label{sec:result} Discovery and exclusion projection}
Next, we estimate the discovery and exclusion projection for different $m_{H^{\pm\pm}}$. Following the Refs.~\cite{Cowan:2010js,Li:1983fv,Cousins:2007yta}, we use the following approximated expressions for the median expected discovery and exclusion significances:

\begin{widetext}

\begin{align*}
Z_{\rm dis} &= \left[ 2\left( (s+b) \ln \left[ \frac{(s+b)(b+\delta_b^2)}{b^2+(s+b)\delta_b^2} \right]-\frac{b^2}{\delta_b^2} \ln\left [1+ \frac{\delta_b^2 s}{b(b+\delta_b^2)} \right] \right) \right]^{1/2},
\\
Z_{\rm exc} &= \left[ 2 \left\{ s-b \ln \left( \frac{b+s+x}{2b} \right) - \frac{b^2}{\delta_b^2} \ln \left( \frac{b-s+x}{2b} \right) \right\} - (b+s-x)(1+b/\delta_b^2) \right]^{1/2}
\label{eq:zdis2},
\end{align*}
where $x = \sqrt{(s+b)^2 - 4sb\delta_b^2/(b+\delta_b^2)}$, $s$ and $b$ are number of signal and background events, respectively, and $\delta_b$ is the uncertainty in the measurement of the background. 

\begin{figure}[htb!]
\centering
\includegraphics[width=0.24\columnwidth]{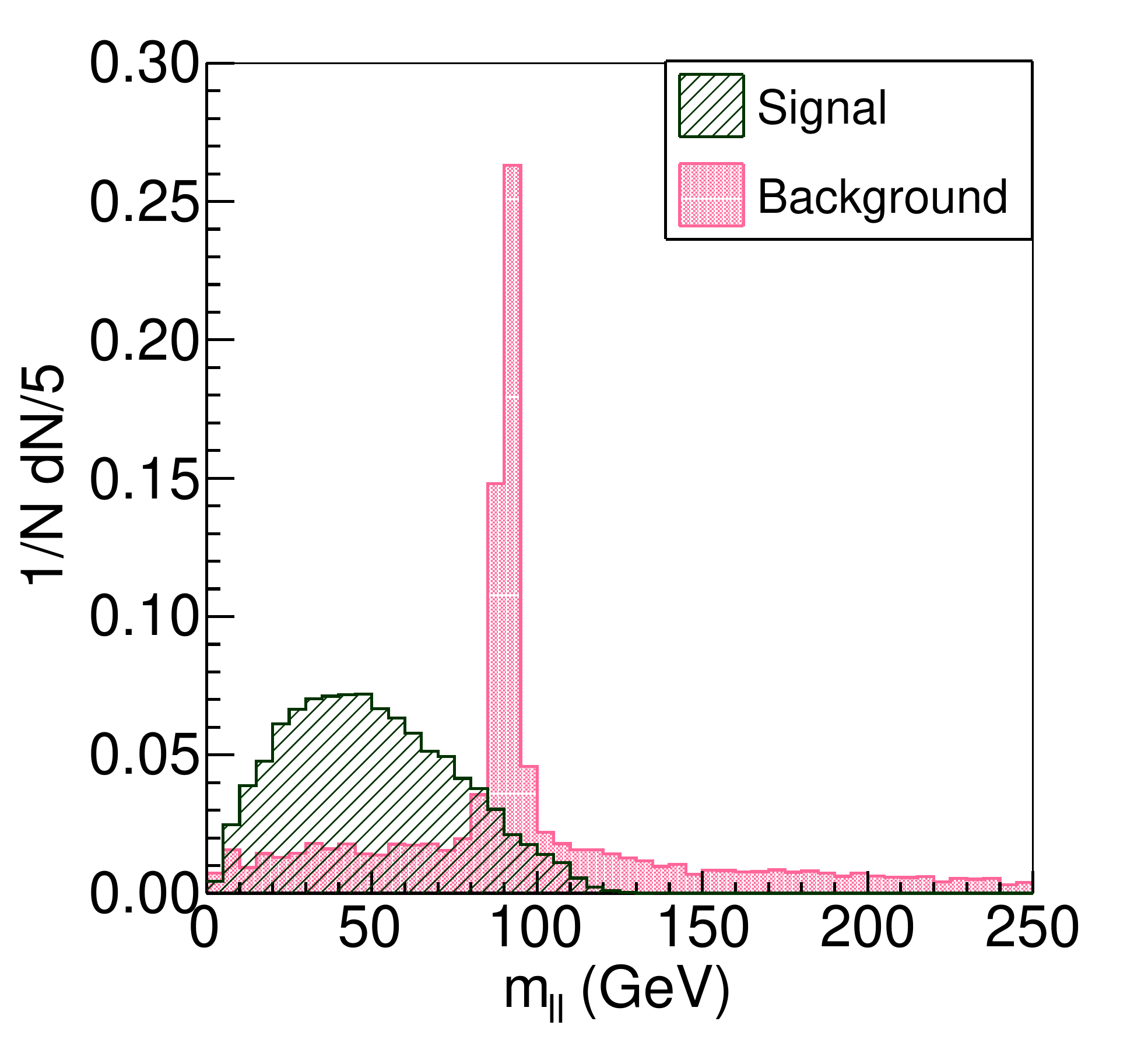}
\includegraphics[width=0.24\columnwidth]{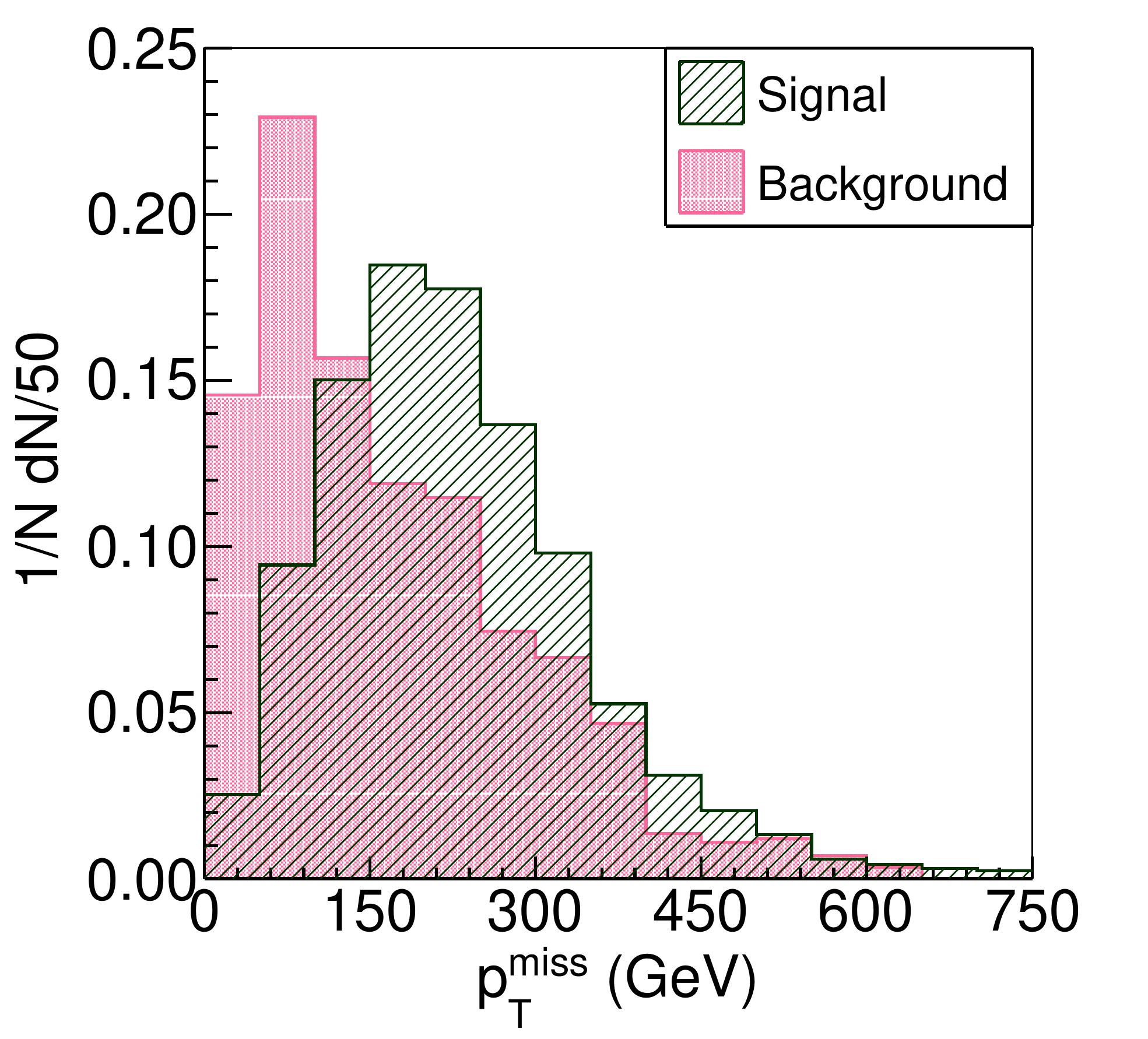}
\includegraphics[width=0.24\columnwidth]{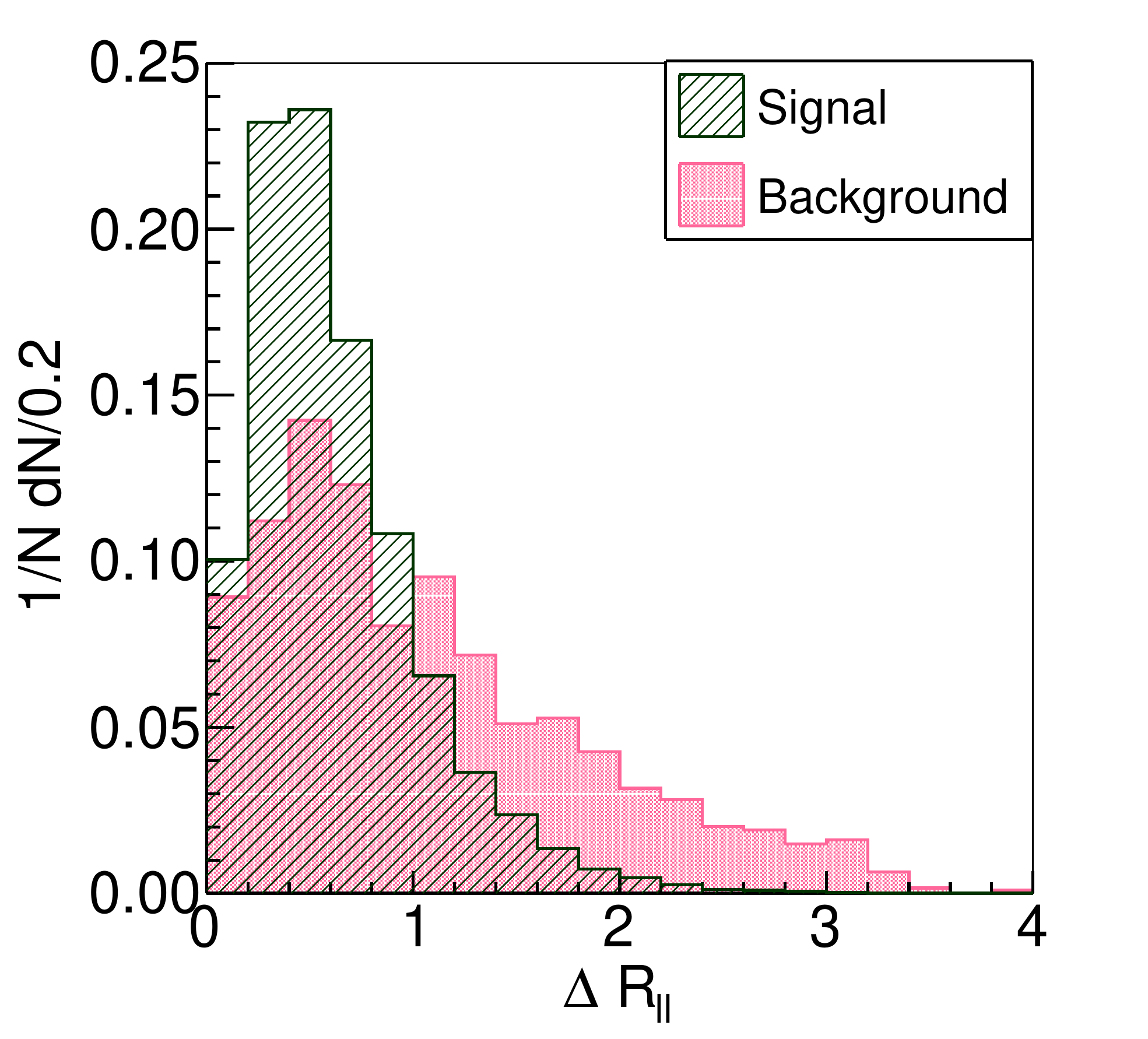}
\includegraphics[width=0.24\columnwidth]{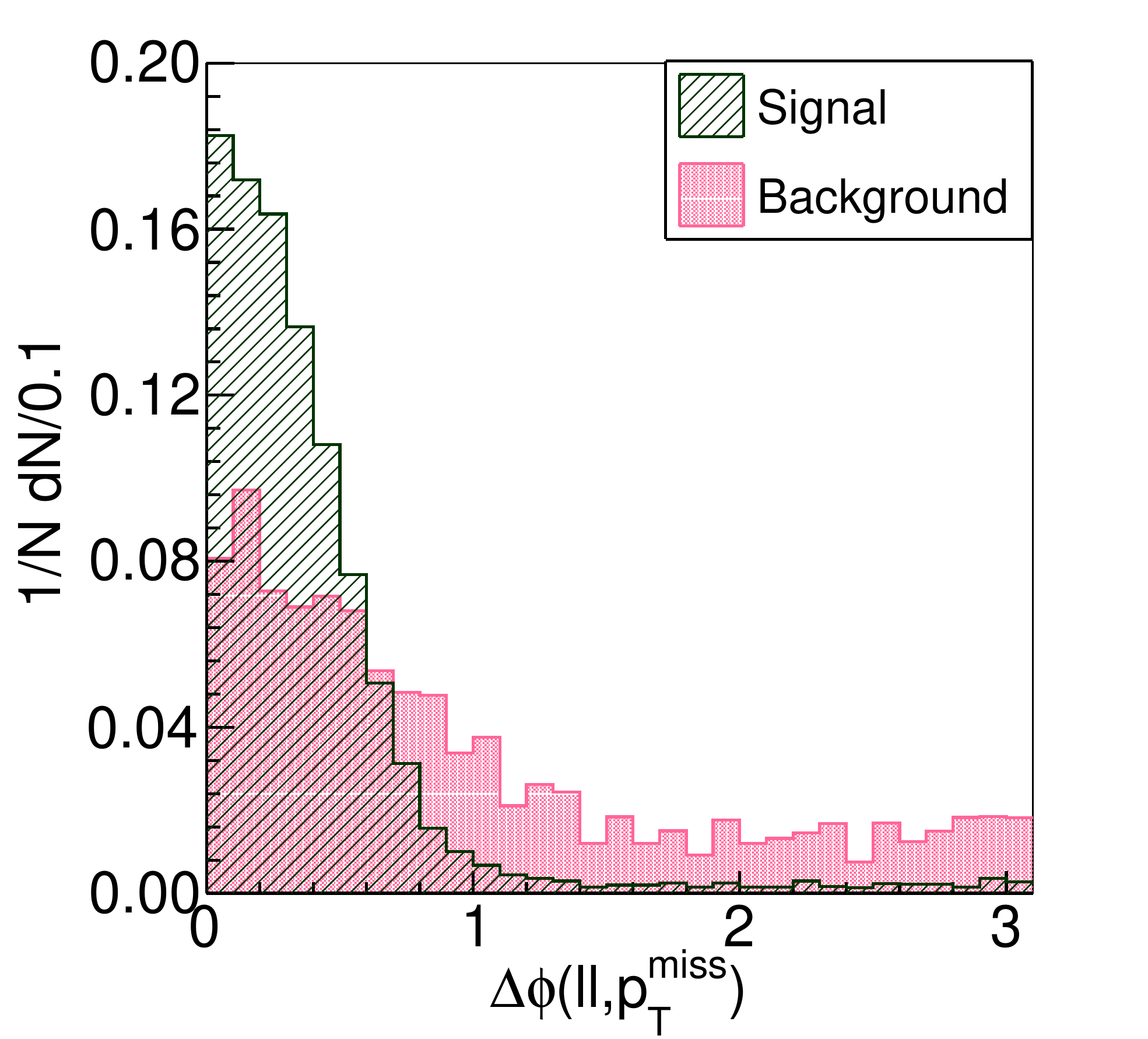}
\caption{\label{fig:mlldRll} Normalised distribution for the signal with $m_{H^{\pm\pm}}=150$ GeV and background events. (From the left to right) First: $m_{\ell\ell}$ after the {\it S1} cut; second, third and fourth: $p_T^{\rm miss}$, $\Delta R_{\ell\ell}$ and $\Delta \phi (\ell \ell, p_T^{\rm miss})$, respectively, after the {\it S2} cut.}
\end{figure}

\end{widetext}

The estimation of the background uncertainty arising from several sources such as the reconstruction, identification, isolation and trigger efficiency, the energy scale and resolution of different physics objects, the luminosity measurements, the pile-up modelling, the parton-shower modelling, the higher-order QCD corrections, {\it etc.} is beyond the scope of this work. We adopt a conservative approach, following the typical LHC searches \cite{CMS:2019lwf,ATLAS:2021eyc}, for which both the theoretical and experimental uncertainties are O(10)\% each, we assume an overall 20\% total uncertainty for the same.

\begin{figure}[htb!]
\centering
\includegraphics[width=0.99\columnwidth]{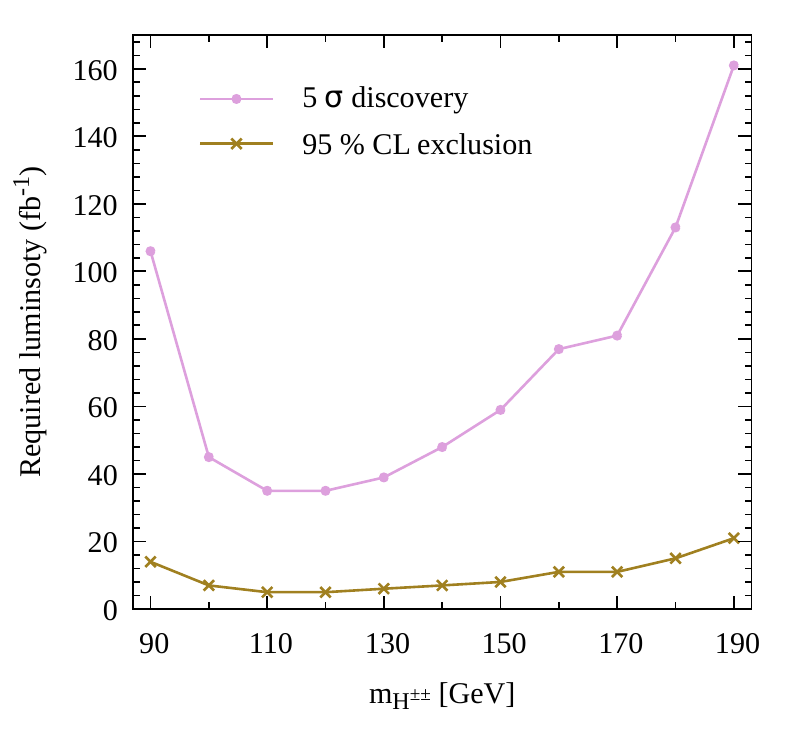}
\caption{\label{table:result} Required luminosity (fb$^{-1}$) for the $5\sigma$ discovery and 95\% exclusion for different $m_{H^{\pm\pm}}$.}
\end{figure}

In Table~\ref{table:result}, we show the required luminosities (in fb$^{-1}$) needed to achieve a median expected $Z_{\rm exc} \geq 1.645$ (95\% CL exclusion) as well as $Z_{\rm dis} \geq 5$ ($5\sigma$ discovery) for different $m_{H^{\pm\pm}}$. The rise in the required luminosity for $m_{H^{\pm\pm}} \lesssim 100$ GeV could be attributed to, as discussed in the end of Section~\ref{sec:bdt}, the poor separation between the $H^{\pm\pm}$-jets and the SM jets, wheras that for larger masses is due to the fall in the signal cross-section (see Fig.~\ref{fig:xsec}).

We find that $H^{\pm\pm}$ within the [84,200] GeV mass range could be probed with $5\sigma$ discovery significane with the already collected Run 2 LHC data. On the other hand, in the case of the data found to be consistent with the SM background, only a fraction of the collected data suffices to exclude them at the 95\% CL.

%\begin{table}[htb!]
%\begin{center}
%\begin{tabular}{|c|c|c|c|c|c|c|c|c|c|c|c|c|c|}
%\hline 
%$m_{H^{\pm\pm}}$ (GeV) & 90 & 100 & 110 & 120 & 130 & 140 & 150 & 160 & 170 & 180 & 190 \\ 
%\hline 
%$5\sigma$ discovery & 106 & 45 & 35 & 35 & 39 & 48 & 59 & 77 & 81 & 113 & 161 \\ 
%95\% CL exclusion & 14 & 7 & 5 & 5 & 6 & 7 & 8 & 11 & 11 & 15 & 21 \\ 
%\hline 
%\end{tabular} 
%\caption{\label{table:result} Required luminosity (fb$^{-1}$) for the $5\sigma$ discovery and 95\% exclusion for different $m_{H^{\pm\pm}}$.}
%\end{center}
%\end{table}

\section{\label{sec:conclusion} Summary}
Doubly-charged Higgs bosons within the mass range 84–200 GeV decaying into a pair of W-bosons have been overlooked by the LHC searches. Lately, Refs.~\cite{Kanemura:2022ahw,Heeck:2022fvl,Bahl:2022gqg,Cheng:2022hbo} have demonstrated that the recently reported measurement of the $W$-bosoon mass by the CDF experiment can be accomodated within the type-II see-saw model predicting such low-mass $H^{\pm \pm}$ and slightly heavier singly-charged and neutral scalars. In view of this, it has been paramount to look for such $H^{\pm \pm}$ at the LHC. In this work, we have presented a novel search strategy for such $H^{\pm \pm}$ considering their pair production in a highly Lorentz-boosted regime such that they are produced back-to-back with large transverse momenta, manifesting themselves as a single fat-jet or a pair of adjacent same-sign leptons plus missing transverse momentum. First, we perform a multivariate analysis to discern such exotic $H^{\pm\pm}$-jets from the SM jets. Then, we perform a search in the final state with an $H^{\pm\pm}$-jet and two same-sign leptons plus missing transverse momentum. We find that such low-mass $H^{\pm\pm}$ could be directly probed with the already collected Run 2 LHC data.

In closing this section, we mention that the search strategy presented here is applicable to any low-mass BSM Higgses (charged as well as neutral) decaying into a pair of SM gauge bosons.

\acknowledgments SA acknowledges the SERB Core Research Grant CRG/2018/004889, and KG acknowledges the DST INSPIRE Research Grant DST/INSPIRE/04/2014/002158 and SERB Core Research Grant CRG/2019/006831. The simulations were supported in part by the SAMKHYA: High Performance Computing Facility provided by Institute of Physics, Bhubaneswar.

\vspace{0.3cm}

\noindent {\bf Note added:} While preparing this manuscript, an article \cite{Butterworth:2022dkt} with similar motivation appeared on the arXiv, concluding that the most of the favoured space for the CDF discrepancy is already excluded by the existing LHC Run 2 data. While our proposed search strategy is completely different from Ref.~\cite{Butterworth:2022dkt}, we also arrived at the same conclusion, i.e., the LHC run II data is sufficient to probe the low mass doubly charged Higgs bosons in type-II seesaw model. Moreover, our strategy is applicable to any low-mass BSM Higgses (charged as well as neutral) decaying into a pair of SM gauge bosons.

\bibliography{v0}

\end{document}